\documentclass[acmtog]{acmart}
\AtBeginDocument{%
  }

\setcopyright{acmlicensed}
\copyrightyear{2025}
\acmYear{2025}
\acmDOI{10.1145/3721238.3730690}
\acmConference[SIGGRAPH Conference Papers' 25]{}
  {August 10-14, 2025}{Vancouver, BC, Canada}
\acmISBN{979-8-4007-1540-2/2025/08}

\PassOptionsToPackage{svgnames}{xcolor}
\usepackage{xcolor}
\usepackage{wrapfig}  

\usepackage{multirow}
\usepackage{slashbox}
\usepackage{booktabs} 

\usepackage{ragged2e}
\usepackage[normalem]{ulem}
\usepackage{cleveref}
\usepackage{bm}
\usepackage{makecell} 
\usepackage{subfig}

\usepackage{tikz}

\newcommand{\lm}[1]{{\color{black}#1}}

\acmSubmissionID{740}

\begin{document}

\title{Image-Space Collage and Packing
with Differentiable Rendering}

\author{Zhenyu Wang}
\affiliation{%
  \institution{Shenzhen University}
  \country{China}
}

\author{Min Lu*}
\affiliation{%
  \institution{Shenzhen University}
  \country{China}
}

\renewcommand{\shortauthors}{Zhenyu Wang and Min Lu}

\begin{abstract}
Collage and packing techniques are widely used to organize geometric shapes into cohesive visual representations, facilitating the
representation of visual features holistically, as seen in image collages and word clouds. Traditional methods often rely on object-space optimization, requiring intricate geometric descriptors and energy functions to handle complex shapes. In this paper, we introduce a versatile image-space collage technique. Leveraging a differentiable renderer, our method effectively optimizes the object layout with image-space losses, bringing the benefit of fixed complexity and easy accommodation of various shapes. 
Applying a hierarchical resolution strategy in image space, our method efficiently optimizes the collage with fast convergence, large coarse steps first and then small precise steps. The diverse visual expressiveness of our approach is demonstrated through various examples. Experimental results show that our method achieves an order of magnitude speedup performance compared to state-of-the-art techniques.
\end{abstract}

\ccsdesc[500]{Computing methodologies~Shape analysis}
\ccsdesc[500]{Computing methodologies~Image-based rendering}

\keywords{Collage, Differetiable Rendering, Image Space}

\begin{teaserfigure}
  \includegraphics[width=\textwidth]{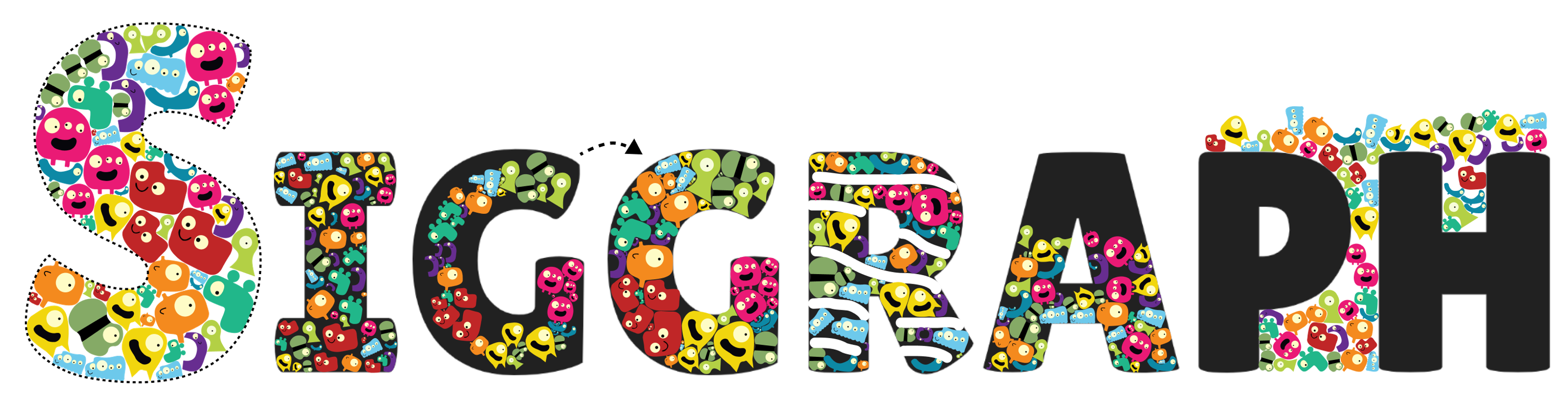}
  \caption{The ‘SIGGRAPH’ example is created using our method: (S) demonstrates the fundamental approach with good shape containment, non-overlap, and uniform distribution; (I) incorporates padding around geometric elements; (G-G) illustrates the smooth transition from axis-based initialization to final filling; (R) showcases collages with stripe blocks; (A) highlights packing within a shape with a downward force; and (P, H) display open packing arrangement with a downward force.}
  \label{fig:teaser}
\end{teaserfigure}


\maketitle

 \section{Introduction}


Assembling and collaging geometric elements to encapsulate visual features provide a unified representation, which has been instrumental in creating intriguing visual designs and artworks, such as circular packing maps to show the thematic topic~\cite{dorling2011}, word clouds for engaging overview of texts~\cite{ramsden2008using}, or digital collages of photos~\cite{spielmann1999aesthetic}. Despite its popularity across various fields, the task of packing elements into given regions presents significant challenges. Numerous techniques have been proposed to address this task, with the majority of existing collage methodologies concentrating on \textit{object-space} optimization~\cite{kwan2016pyramid, wang2019shapewordle, minarvcik2024minkowski, saputra2019improved}. In object space, measuring the fit between geometric objects often involves designing geometric descriptors and energy functions specifically tailored to address the complexity of the objects' shapes.


\begin{wrapfigure}{r}{0.25\textwidth} 
    \centering
    \includegraphics[width=0.25\textwidth]{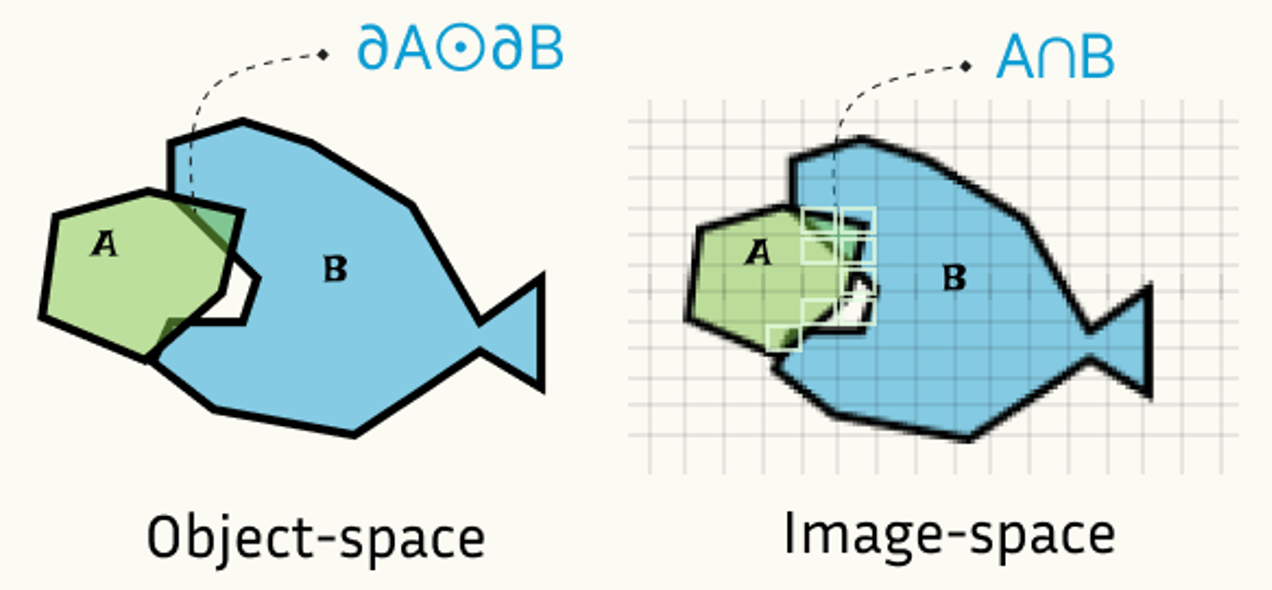} 
\end{wrapfigure} 
Object-based techniques frame collages as a geometric constraint satisfaction problem, accompanied by certain limitations. First, geometric shapes usually need careful analysis to enable effective shape matching~\cite{van2011survey, kwan2016pyramid}. For instance, reducing the overlap between shapes $A$ and $B$ necessitates the shape descriptors for their boundaries ($\partial A$ and $\partial B$). Additionally, geometric descriptors often lack generalizability. For instance, some works necessitate shapes with curvature and are unable to handle open shapes~\cite{kwan2016pyramid}. Some others are limited to fitting containers within convex boundaries~\cite{wang2019shapewordle}. Furthermore, the optimization process in object-based approaches can be computationally intensive, depending on the scale and complexity of the objects involved.

In this work, we advocate a paradigm shift in shape collage techniques by transitioning the geometric packing optimization from the object space to the image space. The core idea is to cast the geometric representation and their spatial relationships onto a grid of pixels, which are with fixed and object-independent complexity. Leveraging the power of differentiable rendering~\cite{li2020differentiable}, our method enables gradient backpropagation from image-space losses to geometric objects, effectively steering the collage optimization process from the discrete image space. Operating in image space facilitates a hierarchical resolution approach to dynamically manage the precision of these image-space losses. The collage process begins with low-resolution losses to facilitate large, bold adjustments and progressively increases resolution for finer refinements. This hierarchical approach significantly accelerates the computation, achieving an order-of-magnitude speedup compared to state-of-the-art methods.

The key strength of our technique is its ability to bypass complex object problem-solving by leveraging the inherent advantages of image-space optimization. This enables it to fit various geometric shapes into almost any desired target shapes. As shown in Figure~\ref{fig:teaser}, our method demonstrates its versatility by supporting a wide range of design configurations. These range from core space filling, as seen in `S', to packing designs influenced by gravity effects in `A', open-region packing exemplified by `H', and complex shapes, such as the white stripe blocks in `R'. Another advantage of our method, which employs gradual descent, is the smooth animation generated during the collage and packing process, as illustrated by the `G's in Figure~\ref{fig:teaser}. In the evaluation, we compared our collage method against state-of-the-art baselines. Results show that our method significantly surpasses the baselines in visual quality. More importantly, our method achieves a remarkable improvement in computational efficiency and scalability, with gains on the order of magnitude. 








\section{Related Work}

The collage and packing problem have been widely studied~\cite{WASCHER20071109}, including 3D object arrangement~\cite{Hu2016_mosaic, Ma2018_3d, ZHUANG2024103996}. Below, we mainly review the research work related to 2D approaches.


\textbf{Collage} In 2D space, collage can be broadly categorized into two types: geometric graphics and images. Circular packing, exemplified by the work of Wang et al.~\cite{wang2006visualization}, is a common paradigm, with new circles added to the outer periphery of existing ones. Several variations of circular packing have been developed, such as single-axis packing~\cite{liu2015online, zhao2014fluxflow}, generative treemap~\cite{vliegen2006visualizing}, and hierarchical packing strategy~\cite{itoh2004hierarchical}. Irregular shapes have also been considered, with methods like arclength descriptor matching~\cite{kwan2016pyramid} and autocomplete-based optimization~\cite{hsu2020autocomplete}. Saputra et al.~\cite{saputra2019improved} represented objects as mass element meshes and used the repulsion forces between neighboring meshes to even out the negative space. Calligraphic packing, employed for letter composition, has been explored by Xu et al.\cite{xu2007calligraphic} and enhanced for legibility by Zou et al.\cite{zou2016legible}. 
Those collage works deliberately describe primitives with geometric parameters, and then optimize over those parameters. Our approach avoids the need for complex geometric computation and the use of task-specific descriptors within the geometry space. Working in image space, our approach can be easily adapted for a variety of applications.

There is a another bunch of works achieving visually pleasing and balanced packing via a top-bottom manner, which divides the canvas into region cells via tessellation and then adjusts the placement of primitives within the cells. Kim and Pellacini~\cite{kim2002jigsaw} proposed an explicit packing energy
function to optimize tiles for compact layouts. Hiller et al.~\cite{Hiller2003_stippling} optimized the centroid placement of small objects such as dots and lines in the cells. Dalal et al.~\cite{Dalal2006} proposed the Sum of Squared Distance metric for even distribution of the primitives with spatial extent. Further, Reinert et al.~\cite{reinert2013interactive} facilitated real-time computation of the sum of squared distance using GPUs and allowed user customization by example. Unlike these methods with tiles and cell adjustments, our work optimizes primitives without global tessellation constraints. Primitives can overlap initially with overlaps, like the `G' in Figure~\ref{fig:teaser}, and fit into open-shaped containers as shown in the `PH' example of Figure~\ref{fig:teaser}.


Another related research topic is image collection, also called photo collage, which deliberately allows occlusions and blends. Many approaches have been proposed~\cite{wang2006picture, rother2006autocollage}. For example, \lm{ShapeCollage~\cite{shapecollage} supports to interactively make a collage of photos with overlapping among photos.} Rother et al.~\cite{rother2006autocollage} allowed for soft intersection among photos. Goferman et al.~\cite{goferman2010puzzle} fused parts of photos into one whole image. Huang et al.~\cite{huang2011arcimboldo} matched multiple cutouts from the Internet to compose a thematic figure. Liu et al.~\cite{liu2017correlation} extracted salient regions and proposed a correlation-preserved photo collage. Pan et al.~\cite{pan2019content} presented a content-based visual summarization technique for image collections. More recently, instead of matching existing photos, Lee et al.~\cite{lee2023neural} generated collage artwork via reinforcement learning based on a given target image and materials, considering scores such as diversity, aesthetics, etc.

\textbf{Text Filling} Texts can be regarded as special geometric shapes. Considerable research has focused on arranging words to create text-based visual design and word art. Word clouds, popular for visually representing words in a compact layout, have been extensively studied ~\cite{hearst2019evaluation, rivadeneira2007getting, viegas2009participatory}. Tools such as Wordle~\cite{mcnaught2010using} help with the easy creation of word clouds. Cui et al.\cite{cui2010context} proposed a dynamic force-directed model for word cloud layout, which preserves semantic context over time. Wu et al.\cite{wu2011semantic} utilized seam carving to optimize word cloud layouts. Beyond traditional word clouds, researchers have explored filling words within specific shapes. Paulovich et al.~\cite{paulovich2012semantic} introduced a cutting-stock optimization method that optimizes the arrangement of words to maximize space utilization within shapes. ShapeWordle~\cite{wang2019shapewordle} took a different approach by utilizing the Archimedean spiral to accommodate irregular shapes, resulting in visually appealing word cloud compositions. MetroWordle~\cite{li2018metro} combined word clouds with maps, incorporating collision detection for geotags. Chi et al.~\cite{chi2015morphable} presented temporally morphable word cloud technology that allows word clouds to undergo smooth shape transformations over time. Xie et al.~\cite{xie2023creating} proposed animating word cloud for emotional expression.

Some other works support interactive word cloud customization. For example, Koh et al.\cite{koh2010maniwordle} introduced an interactive interface to facilitate user-driven word manipulation within word clouds. Jo et al.~\cite{jo2015wordleplus} introduced WordPlus, which expands the interaction of Wordle by incorporating pen and touch interactions. Additionally, Surazhsky et al.\cite{surazhsky2002artistic} proposed a method for text layout on 3D objects. Maharik et al.\cite{maharik2011digital} used streamline-based techniques to arrange words artistically. Zhang et al.~\cite{zhang2022creating} introduced a word arrangement method that arranges theme-related words at the salient areas. Xu et al.\cite{xu2010structure} introduced a tone-based ASCII art generation method. 

Unlike object arrangement guided by space-filling curves or collision detection, our framework utilizes image-based loss for flexible word fitting, accommodating loose compositions like force-attracted filling in a non-closed constrained boundary.

\section{Preliminaries}


\paragraph{Collage Problem} Given a set of 2D geometric items \( G = \{g_1, g_2, \ldots, g_n\} \), the goal of collaging is to arrange them in a geometric container region \( C \), where each shape \( s_i \) may undergo geometric transformations, including translate (\(\mathbf{t}\)), scale (\(\mathbf{s}\)) and rotate (\(\mathbf{r}\)). The optimization problem is formulated as:

\begin{equation}    
\min_{\mathbf{t}_1, \mathbf{t}_2, \dots, \mathbf{t}_n, \mathbf{r}_1, \mathbf{r}_2, \dots, \mathbf{r}_n, \mathbf{s}_1, \mathbf{s}_2, \dots, \mathbf{s}_n} \mathcal{L}(G, C, \mathbf{t}, \mathbf{r}, \mathbf{s}), 
\end{equation}

where \( \mathcal{L} \) quantifies the arrangement quality within the container \( C \). Non-overlapping and shape containment are two basic constraints: 

\begin{align*}
    & \text{Shape Containment:} \quad G_i(\mathbf{t}_i, \mathbf{r}_i, \mathbf{s}_i) \subseteq C, \quad \forall i = 1, 2, \dots, n \quad \\
    & \text{Non-overlap:} \quad G_i(\mathbf{t}_i, \mathbf{r}_i, \mathbf{s}_i) \cap G_j(\mathbf{t}_j, \mathbf{r}_j, \mathbf{s}_j) = \emptyset, \quad \forall i \neq j  \quad
\end{align*}

\paragraph{Vector Representation} We adopt a uniform vector representation for 2D geometric item of any shape. For each item, a closed area with $N$ cubic Bézier curves $\{(p_x, p_y)\}_{i=1}^{3N}$ is initialized and fitted to the silhouette of the item via differentiable rendering. The parameter $N$ controls the shape's granularity. In our examples, we set \(N=20\) to provide a balance of efficiency and geometric precision. 


\begin{figure}[!t]
  \centering
    \includegraphics[width=1.\linewidth]{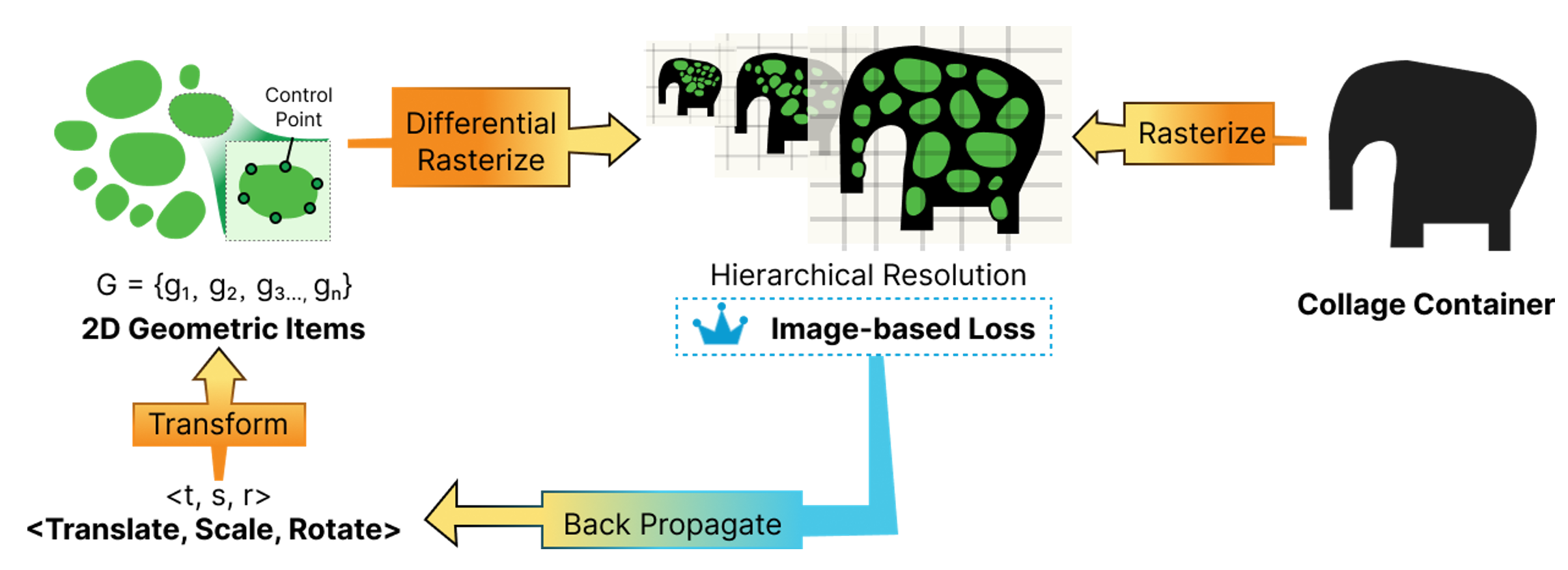}
    \caption{Image-space collage and packing framework: starting with initialized 2D geometric items and their transformations, image-space losses are computed between the rasterized image and the target shape of the collage container across a hierarchy of image resolutions. These losses are then used to iteratively update the transformation parameters, refining the arrangement of the geometric items.}
  \label{fig:overview}
\end{figure}

\paragraph{Differentiable Rendering} Rasterization can be considered as a mapping (or called scene function $I$) from the vector graphics to a 2D pixel grid, denoted as $I(x,y;\Theta)$, where $(x,y)$ is the position of a pixel in the 2D grid, and $\Theta$ represents the vector graphic parameters (e.g., control points of a Bézier curve). Differentiable rendering is a class of techniques that makes the rasterization process differentiable. Differentiable rendering of vector graphics allows the backpropagation from the image domain to the vector graphics domain. Specifically, the scene function $I$ is differentiated with respect to the parameters $\Theta$. Several implementations exist, such as those using differentiable neural networks to approximate rasterization~\cite{zheng2018strokenet, nakano2019neural}. In this work, we adopt the differentiable rendering approach by Li et al.\cite{li2020differentiable}, which leverages the observation that pixel colors become continuous after anti-aliasing. 

\section{Image-space Collage}

Given the collage container $C$, the set of 2D geometric items $G$ are iteratively optimized into a collage, as illustrated in Figure~\ref{fig:overview} for each epoch. First, each geometric item undergoes a geometric transformation applied to its control points $P$, resulting in:

\begin{equation}
\mathbf{P}_{i}' = \mathbf{r}_i \cdot (\mathbf{P}_{i} \odot \mathbf{s}_i) + \mathbf{t}_i, \quad \forall i,
\end{equation}

where geometric items are continuously adjusted by parameters of $\mathbf{t}$ (translation), $\mathbf{s}$ (scaling), and $\mathbf{r}$ (rotation). Geometric items are then rasterized into an image $\hat{I}$ at resolution $w \times h$ by differentiable rendering. The container image is rasterized into a target image $I_C$, where the interior is black and the exterior is white. A series of image-based loss functions are calculated:

\begin{equation}    
\mathcal{L}(\hat{I}(x, y; \mathbf{P}), I_C). 
\end{equation}

Following an update via gradient descent, the image loss is back-propagated to the parameters within the geometric transformation, updated as: 

\begin{equation}    
\mathbf{t}, \mathbf{s}, \mathbf{r} := 
\mathbf{t} - \eta \frac{\partial \mathcal{L}}{\partial \mathbf{t}}, \,
\mathbf{s} - \eta \frac{\partial \mathcal{L}}{\partial \mathbf{s}}, \,
\mathbf{r} - \eta \frac{\partial \mathcal{L}}{\partial \mathbf{r}}.
\end{equation}

Over the optimization process, the image-space loss is calculated among raster images at multiple levels of resolutions, starting with a low resolution and gradually moving to higher resolutions. The transformation parameters are updated iteratively until the arrangement of items converges to an optimal solution.

\subsection{Initialization} 
\label{subsec:init}
\begin{figure}[!htb]
  \centering    \includegraphics[width=.65\linewidth]{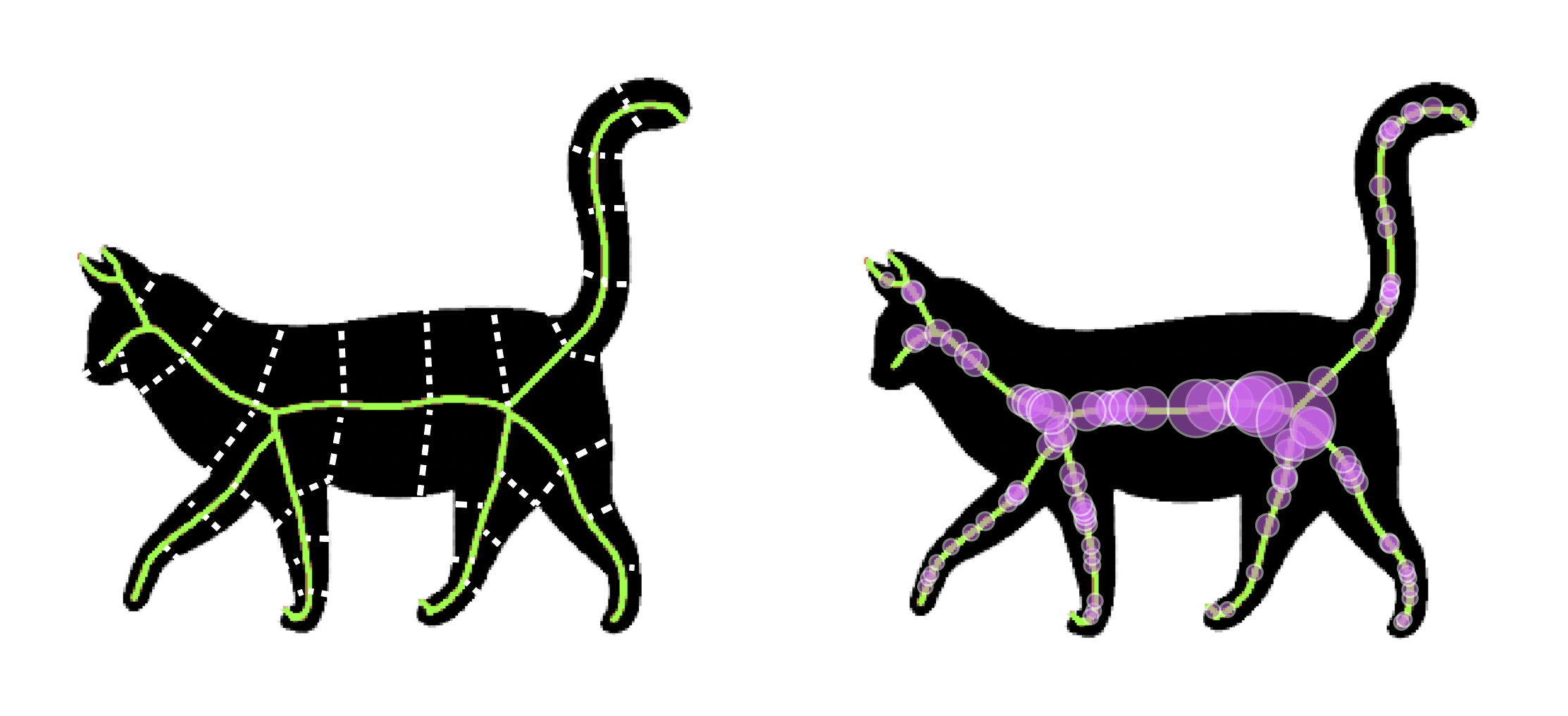}
    \caption{MAT-based position initialization: with the detected medial axes and their nearest associated widths to the boundary (left), visual elements are initialized in the way that larger ones are placed on axes with larger medial widths (right).}
  \label{fig:mat}
\end{figure}

We propose a skeleton-based initialization method for distributing geometric items within the shape container. We use the Medial Axis Transform (MAT)\cite{lee1982medial} to extract the skeleton of the target shape and calculate the medial width (distance to the nearest boundary point). As illustrated in Figure\ref{fig:mat}, visual elements are distributed along the medial axis, with larger elements positioned at points with greater medial width. This approach ensures an even distribution of elements within the shape, making it especially effective for tubular shapes. Note that our method is robust to initialization. In Section~\ref{sec:usage}, we show it can effectively handle poor initialization conditions as discussed in~\cite{minarvcik2024minkowski}.



\subsection{Image-based Loss} 
\label{subsec:loss}

We designed the image-space function to ensure two essential requirements, i.e., shape containment and non-overlapping.



\begin{wrapfigure}{r}{0.12\textwidth} 
    \centering
    \includegraphics[width=0.12\textwidth]{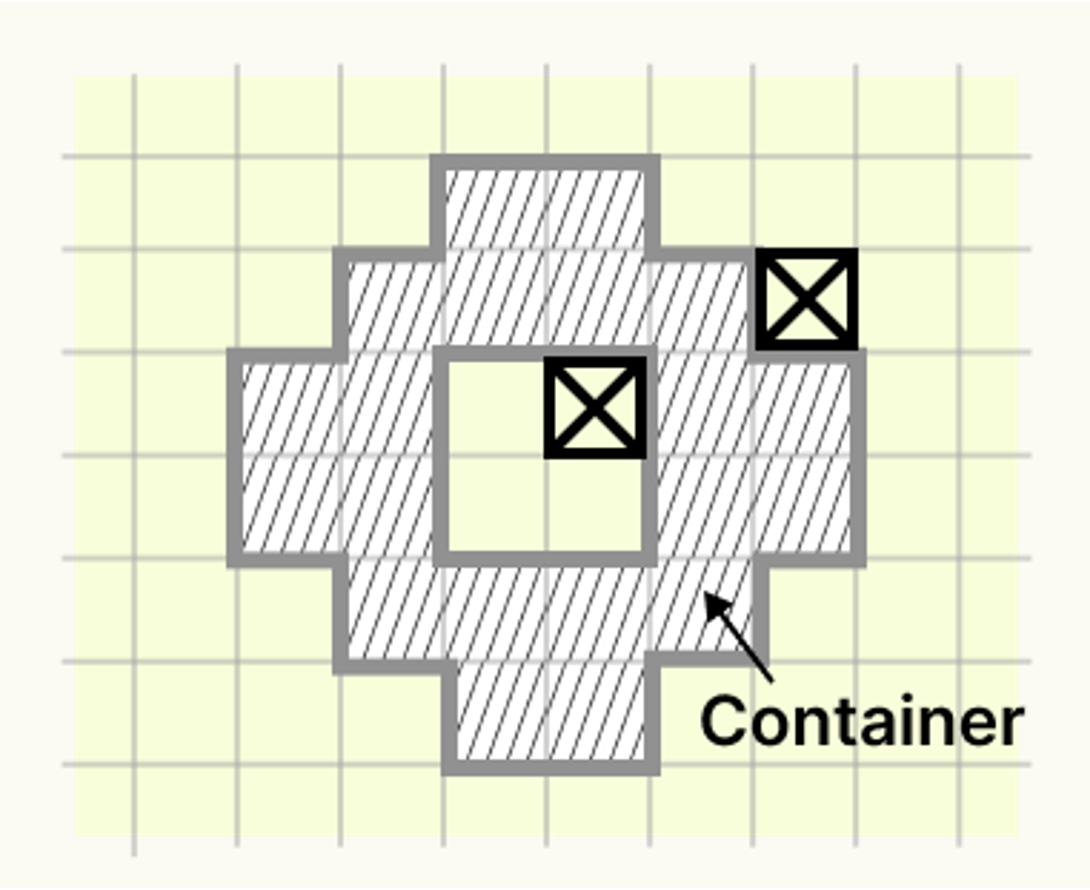} 
\end{wrapfigure}  
\textbf{Shape Containment.} We propose a spatial penalty mask to enhance the basic image mean square error loss (MSE loss) by encouraging elements to full fill the target shape. Geometric items are rendered into black and white image $\hat{I}_{b\&w}$. The penalty mask $W \in \mathbb{R}^{w \times h}$ assigns a small penalty ($w_{ij}=1$) for pixel difference within the target container region and a large penalty ($w_{ij}=100$) for difference outside. The Weighted Mean Square Error (WMSE) between differentiable rasterized image $\hat{I}$ and the target image $I$ is then calculated as:

\begin{equation}
\mathcal{L}_\text{containment}= \frac{1}{w \cdot h} W \odot \left\| \hat{I}_{b\&c} - I_C \right\|^2. 
\label{eq:loss1}
\end{equation}


%

\begin{wrapfigure}{r}{0.12\textwidth} 
    \centering
    \includegraphics[width=0.12\textwidth]{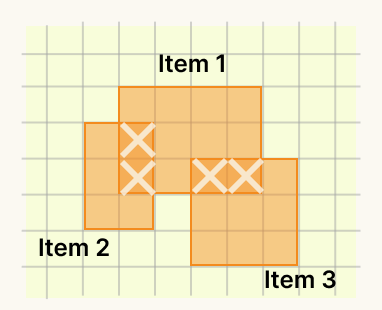} 
\end{wrapfigure} 
\textbf{Non-overlapping.} In image space, detecting overlap among elements is straightforward and avoids geometric computations. Overlap is estimated by rendering all vector primitives with a fixed transparency $\tau$, and and counting pixels whose transparency values deviate from $\tau$, indicating overlapping regions, where \( \mathbb{I} \) is an indicator function for the transparency condition, and $p$ is a pixel of the image: 


\begin{equation}
\mathcal{L}_\text{overlap}=\frac{1}{w \cdot h} \sum_{p} \mathbb{I}(T(p) > \tau). 
\label{eq:loss2}
\end{equation}

\begin{wrapfigure}{r}{0.12\textwidth} 
    \centering
    \includegraphics[width=0.12\textwidth]{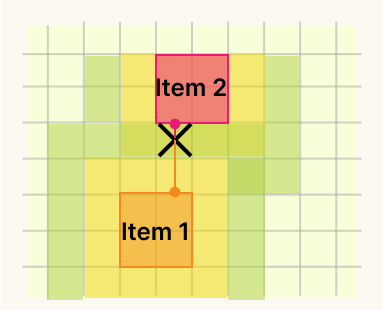} 
\end{wrapfigure} 

\textbf{Even Distribution.} The two loss functions discussed above constrain visual elements within the target shape and prevent overlap, but uneven distribution may still occur, negatively impacting overall visual quality. To address this, we propose a uniform loss $\mathcal{L}_\text{uniform}$. This is achieved through differentiable image dilation ($d$), using a series of convolution kernels with increasing bandwidths (starting at five pixels, incrementing by six pixels per step) to approximate a distance field. As defined in Equation~\ref{eq:lossea}, $\mathcal{L}_\text{uniform}$ is computed as the weighted sum of pixels ($p$) in non-occupied regions within the collage container. The weights ($w$) are assigned based on kernel bandwidths, increasing for larger kernels. Larger dilations highlight broader gaps and are assigned higher weights, while smaller dilations receive lower weights. This approach emphasizes larger spaces, prioritizing their reduction to achieve a more uniform distribution. The equation is as follows:


\lm{
\begin{equation}
\mathcal{L}_\text{uniform}=\sum_{d}\sum_{p} w_d.
\label{eq:lossea}
\end{equation}
}

The overall loss function, shown in Equation~\ref{eqa:all}, uses weights $\alpha$, $\beta$, and $\gamma$ set to 3e3, 8e4, and 5e-4 respectively, to determine the relative contributions of the factors in the optimization process: 

\begin{equation}  
\label{eqa:all}  
\mathcal{L} = \alpha \mathcal{L}_{\text{containment}} + \beta \mathcal{L}_{\text{overlap}} + \gamma \mathcal{L}_{\text{uniform}}.   
\end{equation}


\subsection{Hierarchical Image Resolution}

\begin{figure}[!htb]
  \centering
    \includegraphics[width=1.\linewidth]{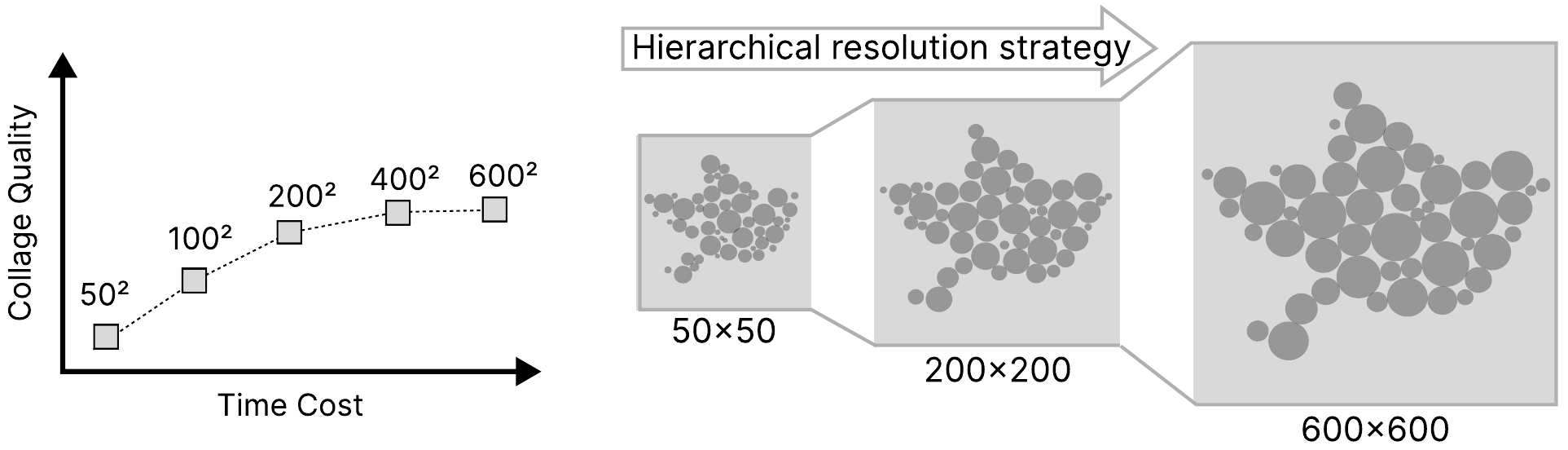}
    \caption{Trade-off between collage quality and computation time for different image resolutions. Note that the three collages on the right have been resized for better visualization of quality differences and do not reflect their original resolution.}
  \label{fig:hie}
\end{figure}

The resolution of the image $\hat{I} \in R^{w \times h}$ plays a crucial role in balancing loss precision and computational efficiency, as is also observed in general image analysis tasks~\cite{Gong2021_multiscale}. Figure~\ref{fig:hie} illustrates this trade-off. Low-resolution images enable faster loss computation but provide lower precision in detecting overlap and containment, resulting in reduced collage quality. Conversely, high-resolution images enhance 
precision and overall collage quality but come with significantly higher computation costs. To balance precision and efficiency, we adopt a hierarchical strategy: starting with a low resolution ($50 \times 50$) to expedite initial computations and progressively increasing to a high resolution ($600 \times 600$) for refinement. Further details are discussed in Section~\ref{subsec:ablation}. 

\section{Results}
\label{sec:usage}

Building on the core image-space collage method introduced earlier, Figure~\ref{fig:gallery0} presents examples of visual collage designs, spanning from intricate vector icons to hand-drawn sketches. Figure~\ref{fig:gallery1} demonstrates how the collage technique integrates seamlessly with images. Below, we explore how this technique can be extended to support a diverse range of use cases.

\begin{figure}[!htb]
  \centering
    \includegraphics[width=1.\linewidth]{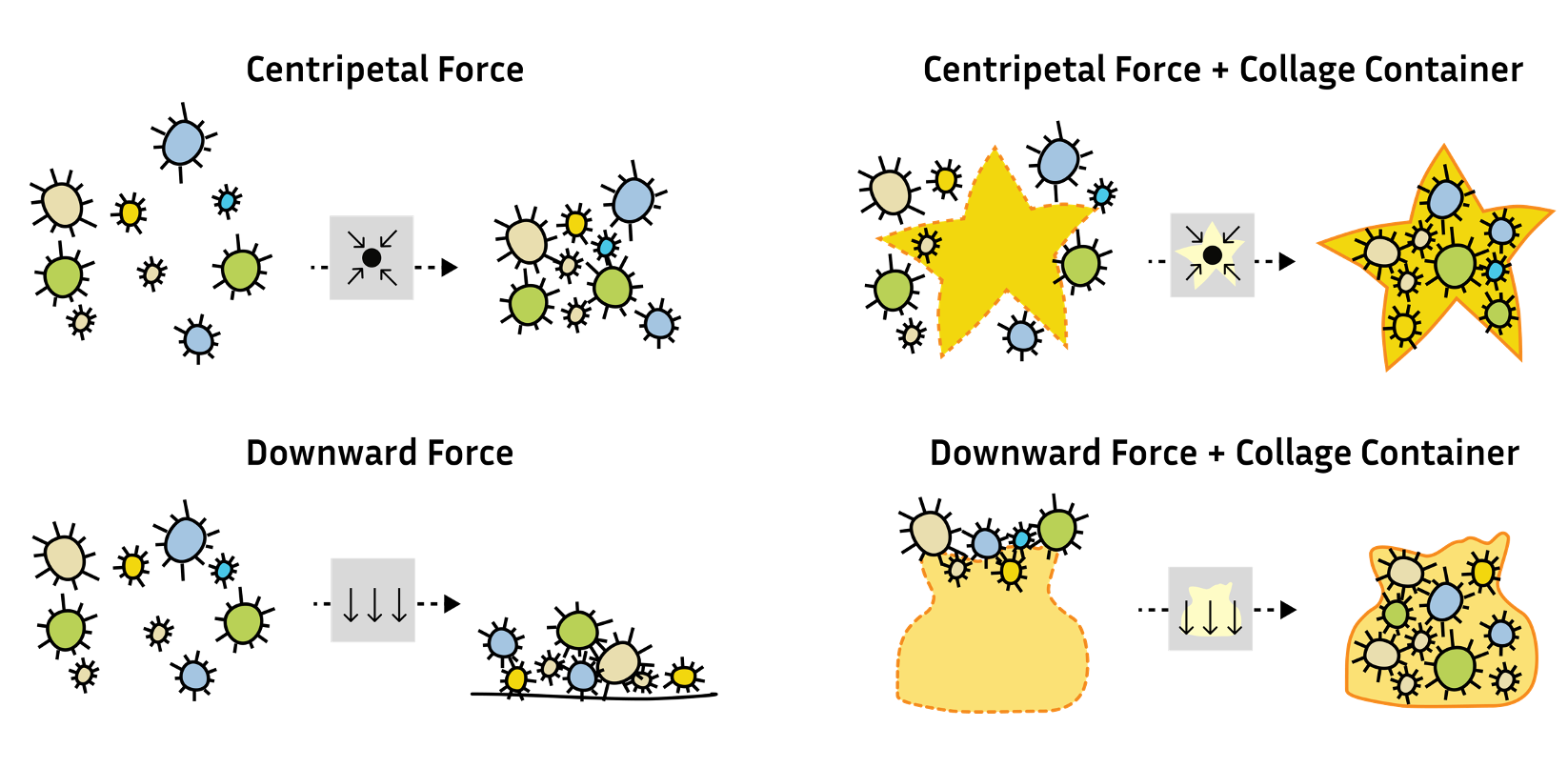}
    \caption{Packing examples with attracting forces: (left) our technique integrates a centripetal or downward force to pack elements efficiently within an open area; (right) using a collage container, elements are first attracted and confined within specific shapes.} 
  \label{fig:forcefield}
\end{figure}

\textbf{Force Attraction} Our method can be seamlessly integrated with force-directed techniques. For example, by defining an attracting or repelling force source, the distance between visual elements and the force source can be computed as a loss function to influence the movement of the elements. This approach enables controlled attraction or repulsion of elements based on the specified force field. As shown in Figure~\ref{fig:forcefield} (left), a packing layout such as a circular or horizontal layout can be achieved using a central force point or a linear downward force. Additionally in Figure~\ref{fig:forcefield} (right), elements can be attracted into the mask by the force attraction with the collage mask. 


\begin{figure}[!tb]
  \centering
    \includegraphics[width=1.\linewidth]{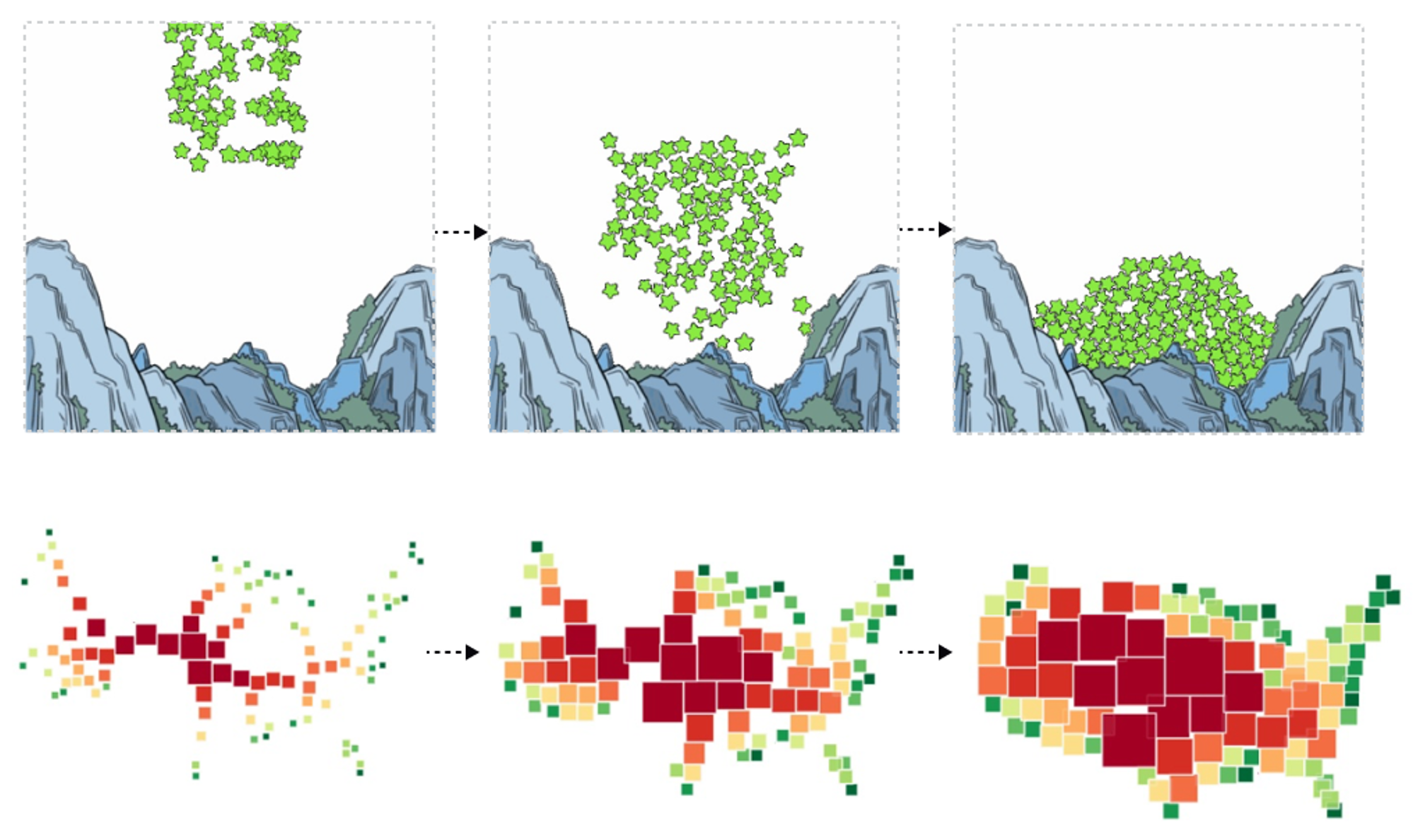}
    \caption{Gradual optimization creates animation effects: (top) an expanding animation effect, (bottom) a falling down animation effect. }
  \label{fig:ani}
\end{figure}

\textbf{Animation Effects} The gradual optimization process of our collage technique produces a side effect: captivating animation effects, distinguishing it from search-and-match algorithms~\cite{kwan2016pyramid}. As shown in Figure~\ref{fig:ani}(top), the star glyphs are initialized on the top line and fall by a downward attracting force, creating an animation effect of `falling down'. In Figure~\ref{fig:ani}(bottom), by the MAT-based initialization, the visual elements move outwards to fit in the shape of the US boundary, creating an animation effect of `expanding'. 


\begin{figure}[!tb]
  \centering
    \includegraphics[width=1.\linewidth]{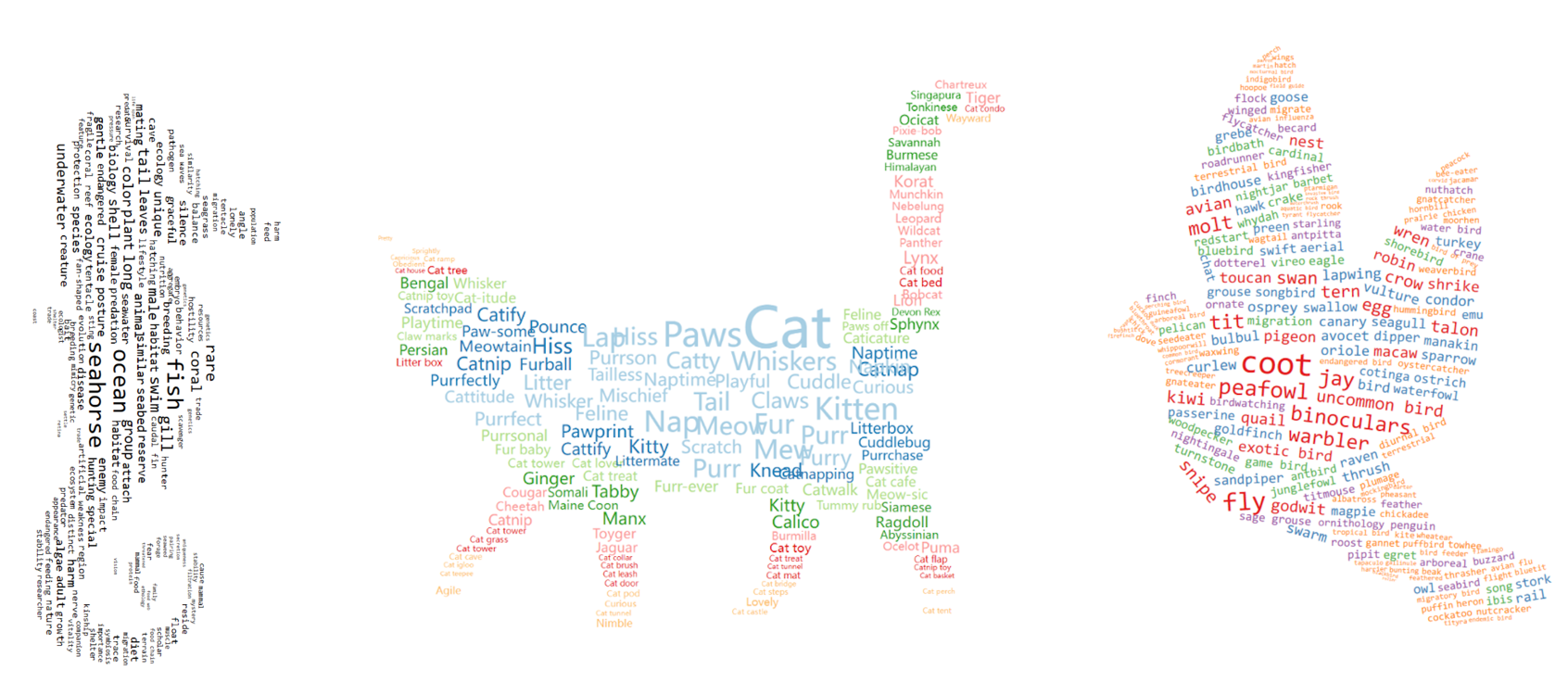}
    \caption{Word clouds that pack words into animal shapes: from left to right, a vertically-aligned collage in the shape of a seahorse, a horizontally-aligned collage in the shape of a cat, and a loosely horizontally-aligned collage in the shape of a bird.}
  \label{fig:wordclouds}
\end{figure}

\textbf{Graphic Text Blending} Our method seamlessly integrates text and graphics, enabling cohesive and visually appealing compositions. As illustrated in Figure~\ref{fig:wordclouds}, our technique is used to generate word clouds (also known as wordles), where words of varying font sizes are arranged to form specific shapes, creating a balanced and engaging design. In Figure~\ref{fig:gallery1}, we show examples of blending texts within image regions with low saliency. This ensures that less prominent areas are utilized effectively, enhancing the overall layout while maintaining the visual emphasis on key elements. 

\begin{figure}[!htb]
  \centering    \includegraphics[width=1.\linewidth]{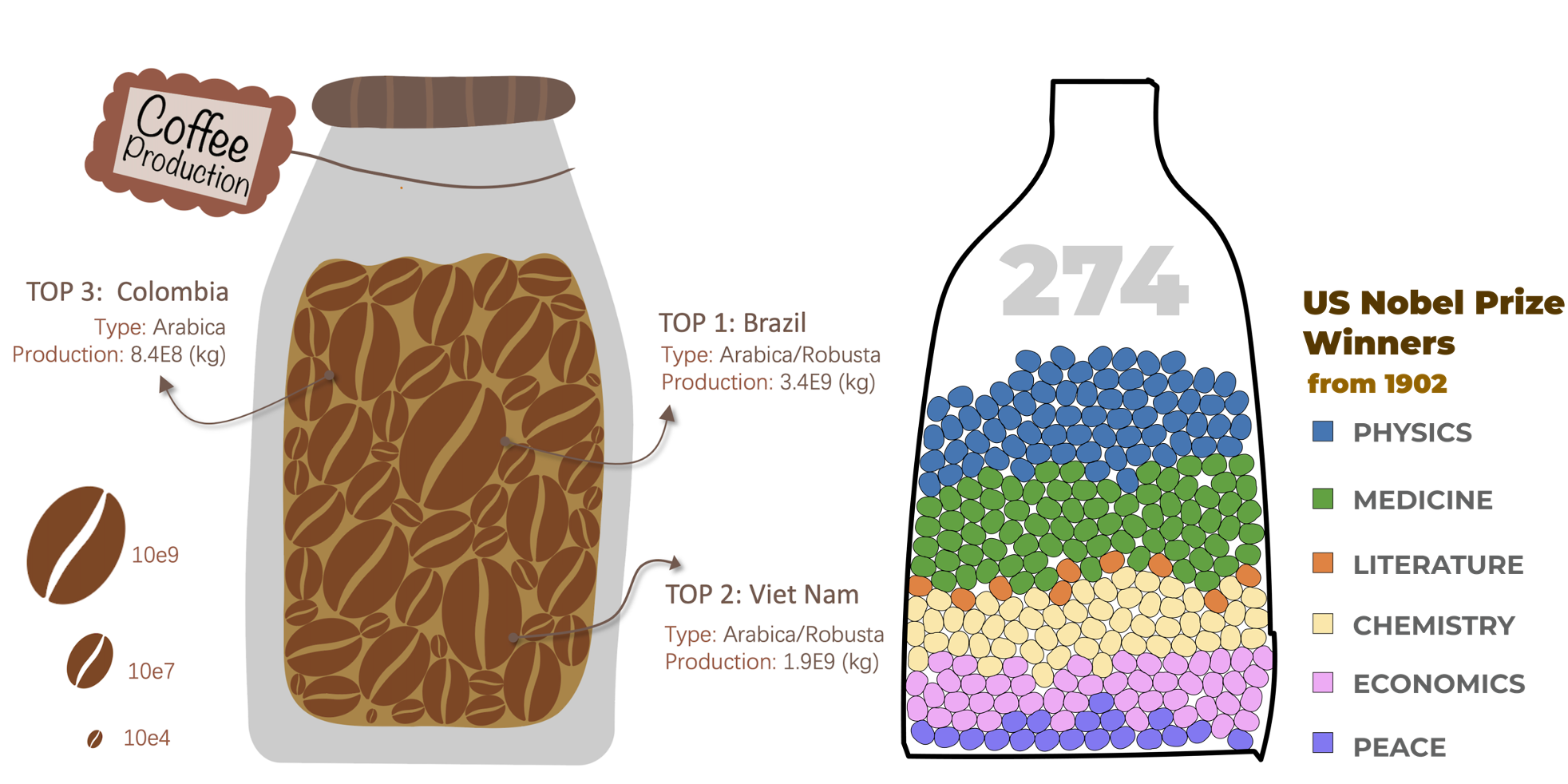}
    \caption{Unit data visualization examples: (left) coffee production infographic, in which each coffee bean is a county that produces coffee, and its size encodes the coffee production. (right) Nobel prize winners in the US, each Nobel winner is represented as a circle, whose color indicates its category.}
  \label{fig:coffee_food}
\end{figure}

\textbf{Data Visualization.} Our method supports unit visualization, where each visual element represents a data item~\cite{park2017atom}. In Figure~\ref{fig:coffee_food} (left), the Country Coffee Production dataset\footnote{https://www.kaggle.com/datasets/michals22/coffee-dataset} is visualized, with each coffee-producing country represented by a coffee bean. The size of the bean encodes the country's coffee production, and the uniform scaling parameter $\mathbf{s}$ ensures accurate area-based encoding without loss of fidelity. Figure~\ref{fig:coffee_food} (right) illustrates a bar-like sedimentary visualization of the Nobel Prize Winners in the US from 1902\footnote{https://www.kaggle.com/datasets/joebeachcapital/nobel-prize}. Each winner is represented by a hand-drawn circle, with colors indicating prize categories, creating a clear and engaging representation of the dataset. In this example, a downward force is integrated to create visual effect of sedimentation. 


\section{Evaluation}


In this section, we first report the results of the ablation study and then elaborate on the comparison between our method and state-of-the-art methods.  

\paragraph{Metrics of Collage Quality} We used three metrics to quantitatively measure the quality of the generated collages. The first metric is adopted from exiting work~\cite{wang2019shapewordle}, and additional two metrics are added to quantify the overlaps among objects and the target shape: 
(1) \textit{Layout Coverage (LC):} it is quantified as the proportion between the number of pixels in the object area (i.e., words in the clouds) and the number of pixels in the non-object area inside the target shape, the bigger the better; (2) \textit{Object Overlap (OO)}: it is quantified as the ratio of pixels in the overlap areas between objects to the total number of pixels in the target shape; (3) \textit{Exceeding Area (EA):} it is quantified as the ratio of pixels that exceed the target shape to the total number of pixels in the target shape. 




\subsection{Ablation Study} 
\label{subsec:ablation}

\paragraph{Ablation on Uniform Loss} We investigated the impact of uniform loss on collage quality, and quantified the \textit{Layout non-Uniformity (L-nU)} as the averaged
distance square of non-object pixels to their nearest objects. As shown in Figure~\ref{fig:uniform}, \textit{without the uniform loss}, the elements are less evenly distributed. For example, in the highlighted regions, collages without uniform loss leave noticeable gaps in other areas, which disrupt the overall balance and aesthetic consistency of the design.

\begin{figure}[!htb]
  \centering
    \includegraphics[width=1.\linewidth]{fig/uniform2.pdf}
    \caption{Comparison of collages with and without the uniform loss: with uniform loss, the elements (e.g., texts on the right example) exhibit a more evenly distributed arrangement.}
  \label{fig:uniform}
\end{figure}

\paragraph{Ablation on Image Resolution Strategies} We conducted an ablation study to evaluate the impact of different image resolution strategies on performance. The study examined eight constant resolution approaches, ranging from 50 to 1200 (shown in Table~\ref{table:vec}), and two hierarchical resolution strategies \emph{100 + 600} and \emph{50 + 200 + 600}.

\begin{figure}[!tb]
  \centering
    \includegraphics[width=1.\linewidth]{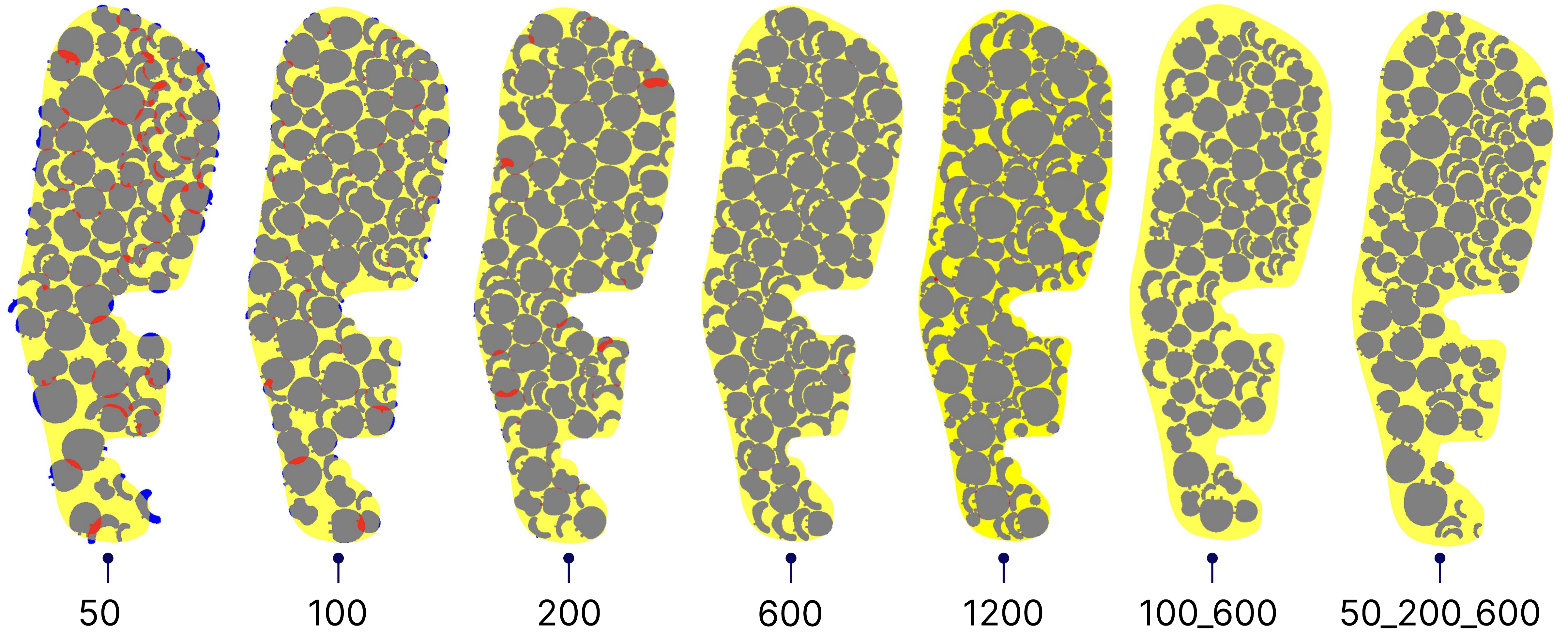}
    \caption{Result samples generated using different resolution strategies: the left five use constant resolutions from low to high, while the right two employ hierarchical resolution strategies.}
  \label{fig:resol}
\end{figure}

The evaluation was conducted under four conditions, packing 100 elements into four different collage shapes to assess the effectiveness of resolution strategies across varying shape complexities. All examples are generated over 200 optimization epochs. For the hierarchical resolution strategy, the epochs were evenly distributed across each resolution level. Figure~\ref{fig:resol} shows collage samples generated using different image resolutions. In the figure, the target shapes are shown in yellow, visual objects in gray, overlaps between visual objects (OO) in red, and areas exceeding the target shapes (EA) in blue.

The averaged results are summarized in Table~\ref{table:vec}, revealing clear trade-offs between resolution strategy and time cost. As can be seen, collage quality improves as the resolution increases. The constant high-resolution $600 \times 600$ achieved high metric scores but incurred significant computational overhead. In contrast, the hierarchical strategies, especially 50 + 200 + 600, demonstrated more balanced performance. They delivered competitive metric results while maintaining lower time costs, highlighting their efficiency in managing resolution adaptively. 

\vspace{10pt} 

\begin{table}[!h]
\small
\caption{Comparison of collage quality and time cost over resolution strategies: time is the cost of 200 optimizing epochs.}
\label{fig:compare_vec}
\centering
\begin{tabular}{c|@{\hskip 6pt}c@{\hskip 6pt}c@{\hskip 6pt}c@{\hskip 10pt}c@{\hskip 6pt}}
\Xhline{1pt}  
    Resolution & Coverage (LC) & Overlap (OO) & Exceed (EA)  & Time (s) \\
        \hline
50x50 & 69.67\% & 2.72\% & 1.11\% & 12.90
\\
100x100 & 65.60\% & 0.68\% & 0.33\% & 13.41
\\
200x200 & 73.43\% & 0.42\% & 0.07\% & 15.44
\\
400x400 & 67.55\% & 0.08\% & 0.01\% & 22.36
\\
600x600 & 70.77\% & 0.11\% & 0.00\% & 32.54 
\\
800x800 & 72.23\% & 0.12\% & 0.00\% & 52.42
\\
1000x1000 & 71.16\% & 0.05\% & 0.00\% & 76.04
\\
1200x1200 & 69.39\% & 0.01\% & 0.00\% & 99.76
\\
\hline
100+600 & 51.47\% & 0.02\% & 0.00\% & 18.51	
\\
\textbf{50+200+600} & 60.73\% & 0.01\%$\downarrow$ & 0.00\% & 20.74 $\downarrow$
\\
\Xhline{1pt}  
\end{tabular}
\label{table:vec}
\end{table}

\subsection{Comparison Experiment}

We compared our method with four existing methods: PAD~\cite{kwan2016pyramid}, Minkowski Penalty~\cite{minarvcik2024minkowski}, ShapeWordle~\cite{wang2019shapewordle}, and ShapeCollage~\cite{shapecollage}. ShapeWordle is designed specifically for text, while ShapeCollage is tailored for rectangular images. Due to the limitations of these methods in handling general geometric shapes, we performed one-on-one comparisons between our approach and each baseline in specific scenarios.

\begin{figure}[!htb]
  \centering
    \includegraphics[width=1.\linewidth]{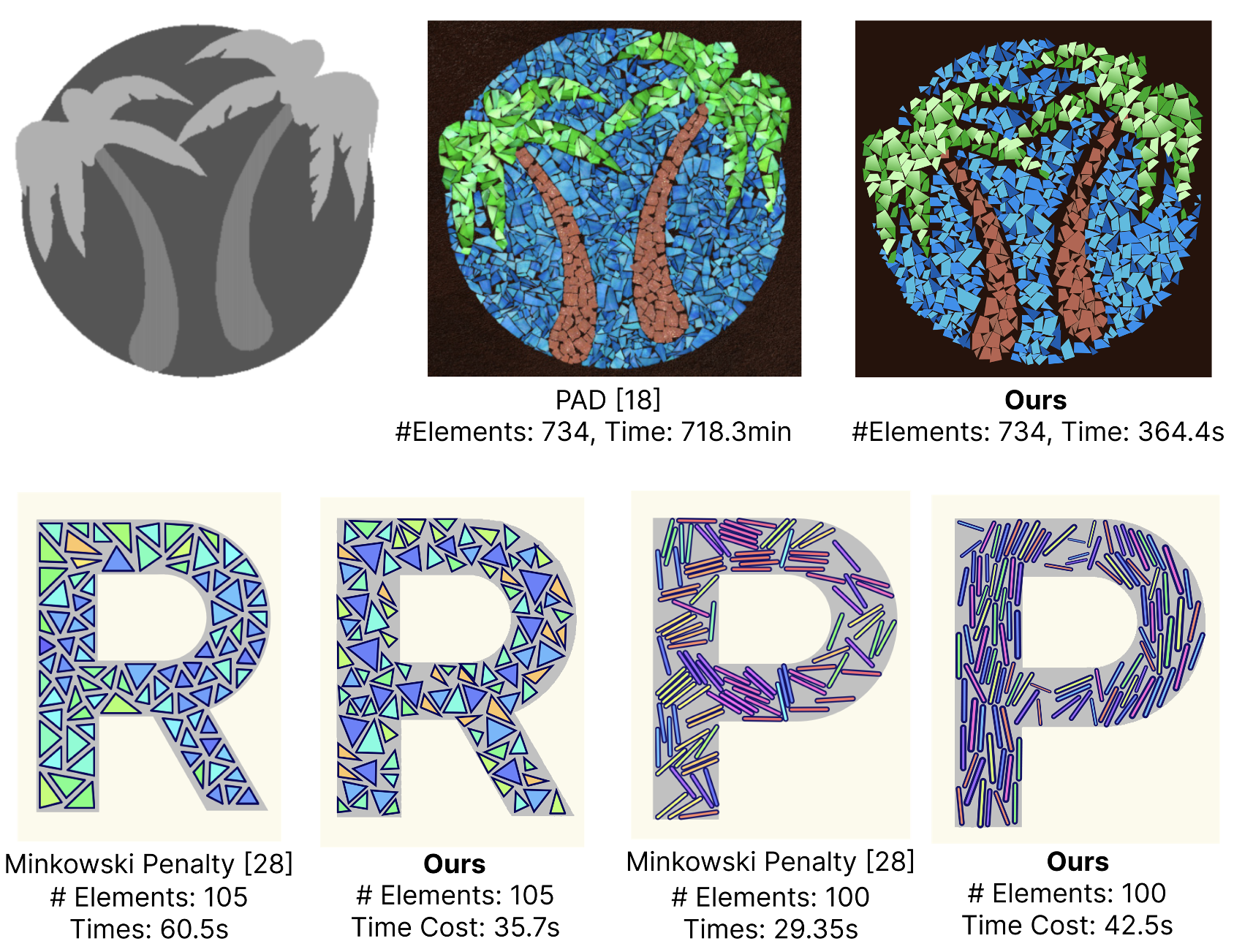}
    \caption{Examples and their time cost generated by our method and others: (top) with PAD~\cite{kwan2016pyramid}, and (bottom) with Minkowski Penalty~\cite{minarvcik2024minkowski}.}
  \label{fig:pad_minski}
\end{figure}

\paragraph{Qualitative Comparison} We compared our method with PAD and Minkowski Penalty, both designed for generating shape collages. As shown in Figure~\ref{fig:pad_minski}, our method delivers visually comparable results to both methods, while being significantly faster. Specifically, our method generates the example in six minutes, compared to over 700 minutes for PAD. Against Minkowski Penalty, our method demonstrates similar time efficiency for the tested examples, around 100 elements. As the number of elements increases, the optimization time for Minkowski Penalty would grow longer due to its $\mathcal{O}(n^2)$ time complexity for arranging constraint pairs.

\begin{figure}[!htb]
  \centering    \includegraphics[width=1.\linewidth]{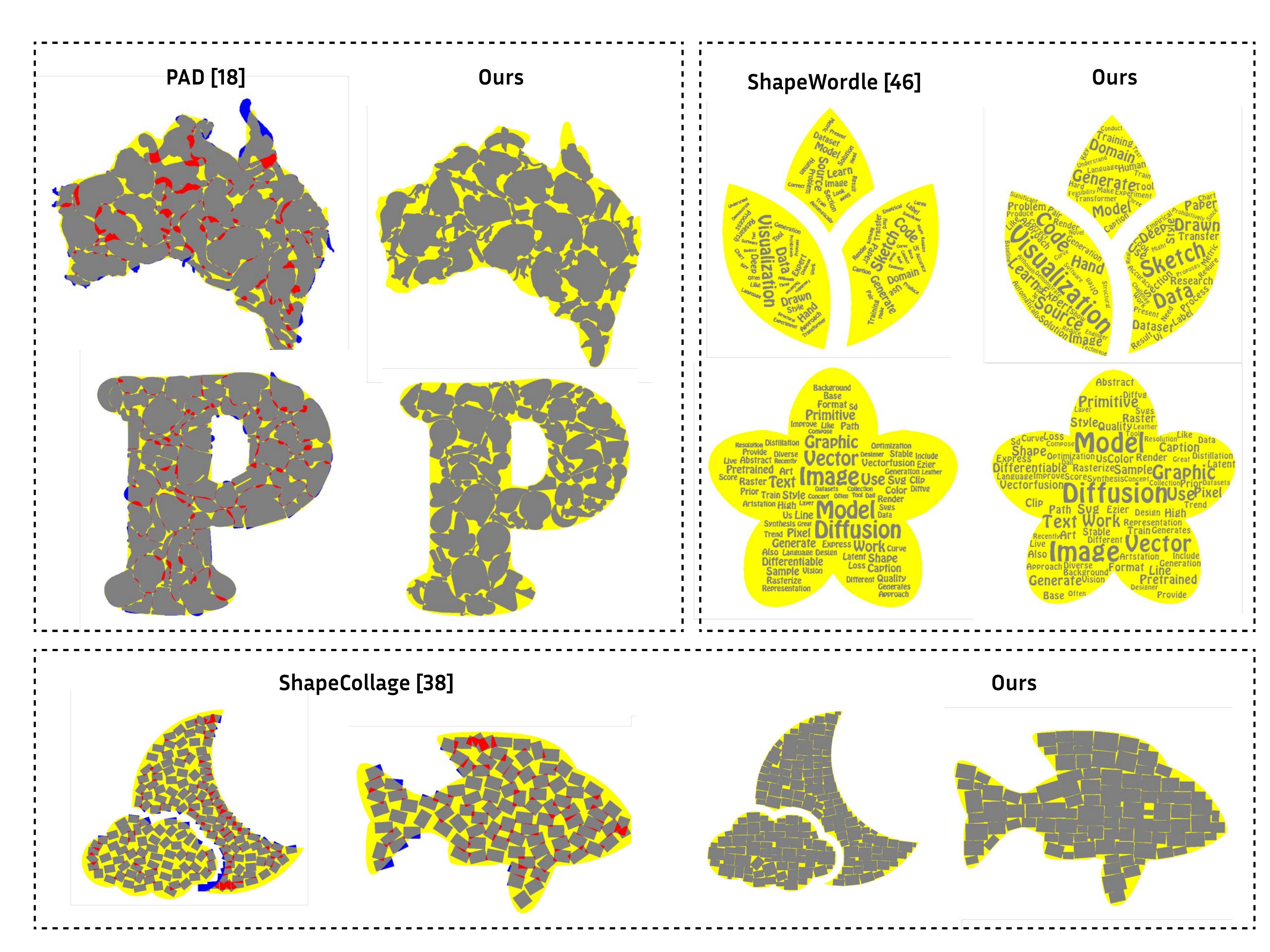}
    \caption{Comparison between our method and three existing methods: yellow areas are the target shape, red areas are where visual objects overlap, and blue areas are where visual objects exceed the target shape.}
  \label{fig:compare}
\end{figure}

\begin{table*}[!htb]
\centering
\caption{\lm{Comparison between our method and three baselines on the six examples in Figure~\ref{fig:compare}.}}
\label{table:baseline}
\small 
\begin{tabular}{c||cc|cc|cc|cc|cc|cc}
\Xhline{1pt}  
\multirow{2}{*}{\backslashbox{{Metrics}}{{Methods}}} & \multicolumn{2}{c|}{{Fig.~\ref{fig:compare} Flower}}             & \multicolumn{2}{c|}{{Fig.~\ref{fig:compare} Leaf}}             & \multicolumn{2}{c|}{{Fig.~\ref{fig:compare} Letter P}}              & \multicolumn{2}{c|}{{Fig.~\ref{fig:compare} Australia}}              & \multicolumn{2}{c|}{Fig.~\ref{fig:compare} Moon}             & \multicolumn{2}{c}{Fig.~\ref{fig:compare} Fish}             \\ \cline{2-13} 
                                  & \multicolumn{1}{c|}{{ShapeW.}} & {Ours} & \multicolumn{1}{c|}{{ShapeW.}} & {Ours} & \multicolumn{1}{c|}{{PAD}} & {Ours} & \multicolumn{1}{c|}{{PAD}} & {Ours} & \multicolumn{1}{c|}{{ShapeC.}} & {Ours} & \multicolumn{1}{c|}{{ShapeC.}} & {Ours} \\ \hline
                            
{Layout Coverage (LC)}                                & \multicolumn{1}{c|}{0.25}   & \textbf{0.28} $\uparrow$     & \multicolumn{1}{c|}{0.18}   &      \textbf{0.29} $\uparrow$         & \multicolumn{1}{c|}{0.94}    &    0.80           & \multicolumn{1}{c|}{0.91}             &   0.75            & \multicolumn{1}{c|}{0.67}            &   \textbf{0.86} $\uparrow$   & \multicolumn{1}{c|}{0.67}            & \textbf{0.87} $\uparrow$      \\ 


{Object Overlap (OO) $ \times 10^{-3}$}                       & \multicolumn{1}{c|}{0}            &       {0}        & \multicolumn{1}{c|}{0}            &     0.02          & \multicolumn{1}{c|}{33.83}             &   \textbf{0.14} $\downarrow$            & \multicolumn{1}{c|}{55.30}             &       \textbf{0.30} $\downarrow$        & \multicolumn{1}{c|}{41.44}            & \textbf{0.09} $\downarrow$     & \multicolumn{1}{c|}{40.04}            &    \textbf{0.08}$\downarrow$  \\
{Exceeding Area (EA)$ \times 10^{-3}$}                       & \multicolumn{1}{c|}{0}            &      0       & \multicolumn{1}{c|}{0}            &    0          & \multicolumn{1}{c|}{7.17}             &    \textbf{0} $\downarrow$        & \multicolumn{1}{c|}{21.84}             &     \textbf{0} $\downarrow $     & \multicolumn{1}{c|}{18.31}            &  0 & \multicolumn{1}{c|}{7.86}            &  \textbf{0}  $\downarrow$ \\ 

\Xhline{1pt}  
\end{tabular}
\end{table*}

\paragraph{Quantitative Comparison} We performed a quantitative comparison between our method and PAD, ShapeWordle and ShapeCollage, based on the three collage quality metrics. We tested ShapeWordle and our method on two target shapes (`flower' and `leaf' in Figure~\ref{fig:compare}) from the ShapeWordle website\footnote{https://www.shapewordle.com/}. PAD and our method were tested using two examples from the PAD work~\cite{kwan2016pyramid} (`Australia' and `Letter P'). For comparison with ShapeImage, two general target shapes were chosen, `Moon' and `Fish'. Table~\ref{table:baseline} summarizes the three metrics of different methods across the six examples.

Figure~\ref{fig:compare} uses the same visual encoding as Figure~\ref{fig:resol}. As can be seen, compared to ShapeWordle, our method consistently demonstrates superior performance in Layout Coverage (LC). In both the `leaf' and `flower' examples, our method has much less space left, and the distribution is more even. Our method outperformed ShapeImage in all three metrics. As can be seen in the `Moon' and `Fish' examples, our method achieves larger coverage, but with much more even distribution, less overlap among objects, and less exceeding from the target shapes. As shown in the `Australia' and `Letter P' examples, PAD gets a more compact layout than ours, with less space left, which results in better scores in Layout Coverage. However, PAD causes more severe overlapping (i.e., the red and blue areas in Figure~\ref{fig:compare}) than ours.

\section{Conclusion and Future Work}

In this work, we have introduced a neat approach to creating collage and packing visualizations by leveraging vector graphics manipulation through an optimization process aimed at minimizing loss in image space. Through the diverse examples presented in Section~\ref{sec:usage}, we have demonstrated the versatility of our method in generating visually compelling collages. Compared to object-based methods such as PAD~\cite{kwan2016pyramid} and Minkowski Penalty~\cite{minarvcik2024minkowski}, our method offers the advantages of being free from object-specific representations and achieving greater computational scalability. Our image-space approach empowers users to explore their creativity, experiment with novel visual elements, and incorporate imaginative concepts, thereby expanding the possibilities for expressive and engaging visual design.

\textit{Link with Image Generation Models} In light of the advancements made, there are several promising directions for future research and development. One potential avenue is the exploration of interactive interfaces for target image-space editing in visualization creation. By providing users with intuitive editing and controls in the target image, they can directly manipulate and refine the visual elements in return, allowing for a more interactive and iterative design process. Designing an interactive system for collage authoring would be an interesting work in the future. A more promising avenue is to import text-driven editing for collage design based on a text-to-image foundation model \cite{iluz2023word}\cite{jain2023vectorfusion}.

\textit{Element Initialization.} In this work, we experimented with one primitive initialization method, MAT-based. It is important to note that different primitive initialization methods can be suitable for different conditions, depending on the specific requirements and constraints of the application. For example, the MAT-based initialization proves effective for shapes with varying widths, such as tubes and necks. As seen in the `tail of seahorse' of  Figure~\ref{fig:wordclouds}, our experiments validate the promising results achieved through the MAT-based initialization technique. However, the MAT-based method is not optimal for target shapes with round bellies. Potential future work is to study adaptive primitive initialization techniques that automatically suggest initial visual primitives based on the geometric features of the target shape. This would enhance the efficiency and accuracy of the initialization process, leading to better adaptation of our method to diverse geometric configurations.

\textit{The Curse of Local Minima.} Like any other iterative optimization algorithms with loss functions, our method can get stuck in some local minima, when visual primitives are not well-fitted in the target shape. When some small primitives are fully contained in some big elements, they are shadow-trapped. This obstruction leads to a state of stagnation, where the primitive remains stationary and unable to move. Some techniques can be used to alleviate the curse of local minima. For example, a sheepherder algorithm can be integrated into the collage optimization procedure, which can monitor and report problems in a global scope, such as detecting the coverage of visual elements, etc.

\textit{Hybrid Object- and Image-space} In Figure~\ref{fig:compare}, we demonstrate that our method outperforms other object-based approaches in collage generation, especially in terms of compactness and non-overlap. However, object-space methods have distinct advantages. For instance, methods like Minkowski Penalty~\cite{minarvcik2024minkowski} provide finer control over object properties, such as preserving balance and harmony among selected objects. A promising future direction would be to incorporate object-space loss into our framework to refine spatial relationships further and enhance layout quality.





\clearpage

\section*{Acknowledgments}

We are deeply grateful to Prof. Daniel Cohen-Or and Prof. Dani Lischinski for their encouragement and insightful feedback throughout this work. We also thank the anonymous reviewers for their constructive suggestions. This work is supported in parts by fundings from Shenzhen Science and Technology Program (20231122121504001), National Natural Science 
Foundation of China (NSFC) Program (62472288), and Guangdong Laboratory of Artificial Intelligence and Digital Economy (SZ), MNR Key Laboratory for Geo-Environmental Monitoring of Great Bay Area, and Guangdong Key Laboratory of Urban Informatics.

\bibliographystyle{ACM-Reference-Format}
\bibliography{template}

\appendix
\clearpage


\begin{figure*}[!thb]
  \centering
    \includegraphics[width=.85\linewidth]{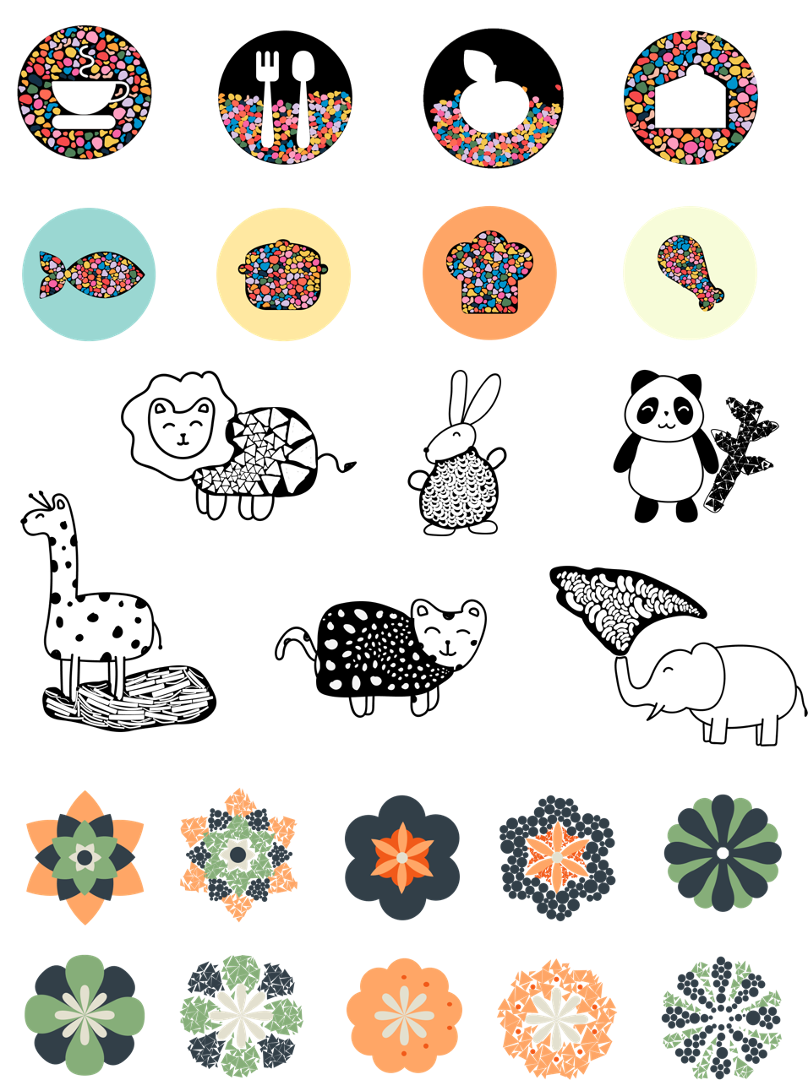}
    \caption{\lm{A gallery of examples: diverse visual elements (i.e., icons, sketched paths) can be effectively fitted within convex and concave target boundaries. }}
  \label{fig:gallery0}
\end{figure*}

\begin{figure*}[!thb]
  \centering
    \includegraphics[width=.8\linewidth]{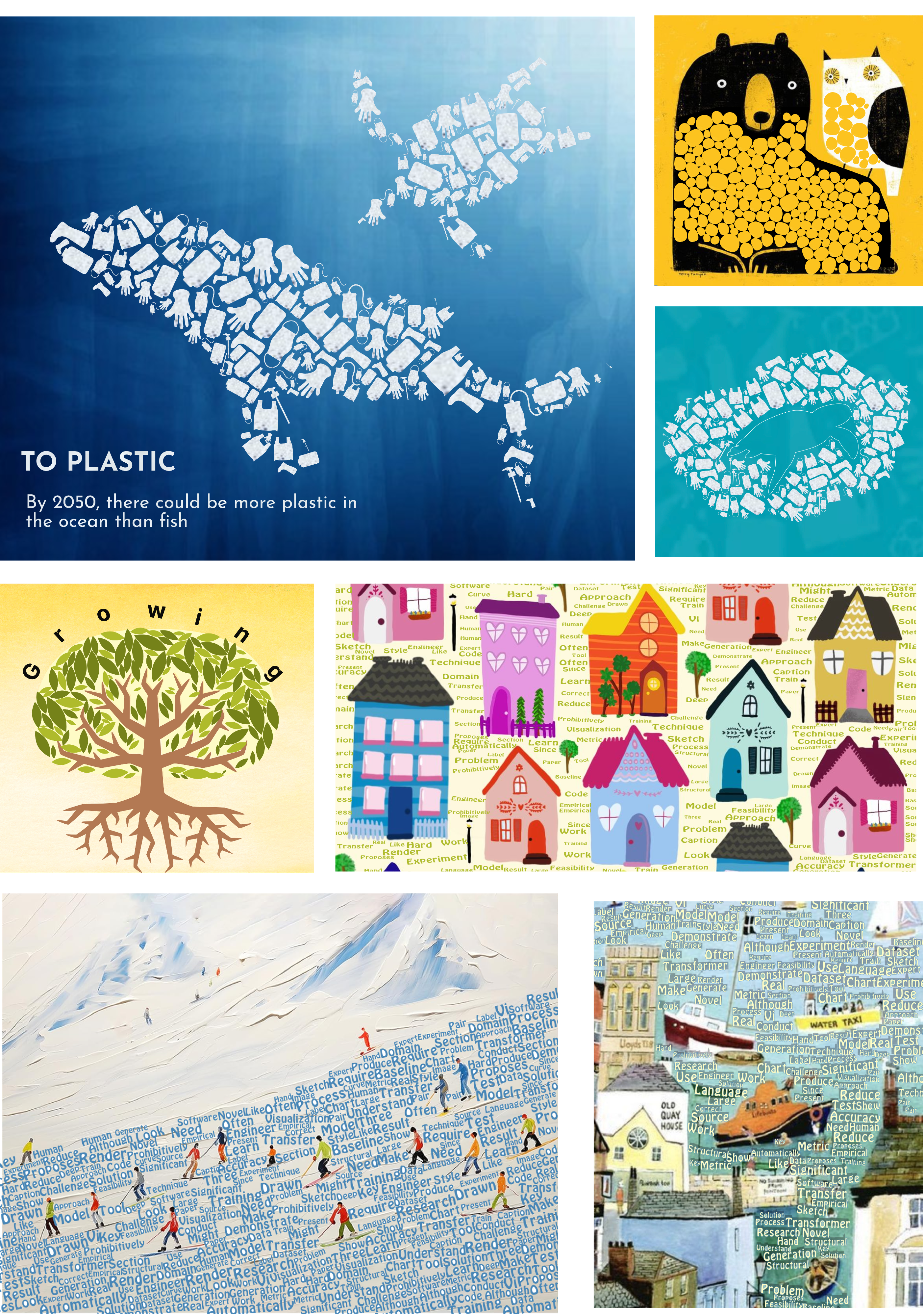}
    \caption{A gallery of examples that texts and graphics are collaged and packed for visually appealing design.}
  \label{fig:gallery1}
\end{figure*}

\end{document}